\documentclass[usenatbib]{aa}
\usepackage{graphicx}	
\usepackage{amsmath}	
\usepackage[colorlinks]{hyperref}
\usepackage{txfonts}
\newcommand{\pmill}{{\sc p-millennium~}}
\newcommand{\pms}{PMS~}

\newcommand{\hmsun}{\, {\rm h}^{-1}{\rm M}_\odot}
\newcommand{\msun}{\, {\rm M}_\odot}

\newcommand{\alfex}{$\alpha_M$ }
\newcommand{\gaea}{{\sc gaea}}
\def\lesssim{\lower.5ex\hbox{$\; \buildrel < \over \sim \;$}}
\def\gtrsim{\lower.5ex\hbox{$\; \buildrel > \over \sim \;$}}
\begin{document}
\title{Reinterpreting the puzzling properties of z>6 galaxies \\ within a variable IMF framework.}
\author{Fabio Fontanot\inst{1,2}\fnmsep\thanks{e-mail:fabio.fontanot@inaf.it},
  Gabriella De Lucia\inst{1,2}, Lizhi Xie\inst{3}, Stefano Zibetti\inst{4}, Francesco La Barbera\inst{5}, \\
  Sebastiano Cantarella\inst{6,1}, Michaela Hirschmann\inst{7,1}, St\'ephane Charlot\inst{8} and Gustavo Bruzual\inst{9}
}
\institute{INAF - Astronomical Observatory of Trieste, via G.B. Tiepolo 11, 
        I-34143 Trieste, Italy
        \and
        IFPU - Institute for Fundamental Physics of the Universe, via Beirut 2,
        34151, Trieste, Italy
        \and
        Tianjin Normal University, Binshuixidao 393, 300387, Tianjin, People’s
        Republic of China
        \and
        INAF–Osservatorio Astrofisico di Arcetri, Largo Enrico Fermi 5, I-50125 Firenze, Italy
        \and
        INAF – Astronomical Observatory of Capodimonte, Sal. Moiariello, 16,
        80131 Napoli, Italy
        \and
        Astronomy Section, Department of Physics, University of Trieste, via G.B. Tiepolo 11, I-34143,
        Trieste, Italy
        \and
        Institute for Physics, Laboratory for Galaxy Evolution, 
        Observatoire de Sauverny, Chemin Pegasi 51, 1290 Versoix, Switzerland
        \and
        Sorbonne Universit\'e, CNRS, UMR 7095, Institut d’Astrophysique de Paris,
        98 bis bd Arago, 75014 Paris, France
        \and
        Instituto de Radioastronom\'\i a y Astrof\'\i sica, UNAM, Campus Morelia,
        C.P. 58089, Morelia, M\'exico
        }

   \date{Received ???, 2026; accepted ???, 2026}

   \abstract{Recent results form the James Webb Space Telescope (JWST)
     report space densities for bright and massive galaxies at z>7
     that far exceed expectations of theoretical models of galaxy
     formation, prompting a revision of our understanding of the
     physical processes leading to the assembly of the first luminous
     structures. In this work we present predictions from a
     realization of the GAlaxy Evolution and Assembly (\gaea) model,
     which implements a prescription for a variable stellar initial
     mass function (IMF). This prescription is inspired by
     high-resolution numerical simulations that account for the role
     of cosmic rays (CR) as regulators of the star formation rate
     (SFR) in giant molecular clouds. In our approach, SFR density is
     assumed to be a proxy for the CR density, providing a link
     between the IMF shape and the predicted physical conditions of
     the star forming interstellar medium. Our results show that, in
     our model framework, assuming such a variable IMF reproduces
     several properties of the z>6 galaxy population, with no further
     modification of the feedback model, including their UV luminosity
     functions up to z$\sim$13. In order to compare model predictions
     with available estimates for the galaxy stellar mass function
     (GSMF), we reconstruct stellar masses from the model's synthetic
     photometry assuming a universal IMF, reflecting standard
     observational practice. Under this approach, we show that the
     model can reproduce the evolution of the GSMF up to the highest
     redshifts accessible. Our findings highlight the need to consider
     a variable IMF shape in the error budget associated with stellar
     mass estimates. We show that the evolution of both the slope and
     normalization of the gas-phase mass metallicity relation can be
     used as powerful discriminant between models of early galaxy
     formation assuming different IMF evolution.}


\titlerunning{High-z Variable IMF}\authorrunning{F.~Fontanot et al.}
   
\maketitle
\nolinenumbers
%
\section{Introduction}
\label{sec:intro}

In recent years, the wealth of data coming from the James Webb Space
Telescope (JWST) opened new challenges in our understanding of the
processes leading to the early formation and assembly of galactic
structures at z$\gtrsim$8-9. The first puzzling evidence came from the
discovery of a large population of UV-bright galaxies at such
redshifts, now accessible thanks to the observed frame IR capabilities
of JWST \citep[see e.g.][]{Robertson22}. Indeed, derived number
densities for this population \citep{PerezGonzalez23, Castellano23,
  Donnan23, Harikane23, Leung23, Donnan24, Harikane24, McLeod24,
  Adams24, Casey24, Finkelstein24, Kokorev25, Whitler25} far exceeded
expected UV luminosity function (UVLF) from earlier generation
theoretical models of galaxy formation. Despite some of the
photometric candidates turned out to be lower-redshift interlopers
\citep[see e.g.][]{ArrabalHaro23}, the number of spectroscopically
confirmed bright z$\gtrsim$10 sources has grown at increasing rate
\citep[among the others]{Carniani24, Harikane24}. Moreover, attempts
to use the available spectro-photometric data to infer the physical
properties of these galaxies, through standard SED-fitting approaches
led to speculations that also their galaxy stellar mass function
\citep[GSMF, see e.g.][]{NavarroCarrera24, Weibel24, Shuntov25,
  Harvey25} provides critical tension with predictions of theoretical
galaxy formation models. Despite the fact that early suggestions of a
tension with the standard $\Lambda$CDM cosmological model
\citep{Labbe23} have been recently toned down \citep[see
  e.g.][]{Yung25}, relevant discrepancies with the predictions of
theoretical models of galaxy formation have been consistently reported
\citep[among the others]{Finkelstein24, Yung24, Cantarella25}. It is
important to keep in mind that most of these models have been
calibrated to reproduce the properties of galaxies at lower redshifts
(typically z$\lesssim$4), using prescriptions to regulate the key
physical mechanism that are also based on observations of the low-z
Universe. Therefore their struggles to reproduce the properties of
higher redshift progenitors can be ascribed to either an
oversimplification of these prescriptions, which are no longer
adequate to treat the different physical properties (in terms of gas
content, metallicity) of high redshift galaxies, or to the lack of
focus on some physical ingredients that are important at early epochs
\citep[see e.g.][]{MaioPeroux26}.

Several explanations have been proposed to explain the tension between
theoretical predictions and the observed UVLF and GSMF at
z$\gtrsim$7-8. From the observational point of view, the estimated
relevant number of Active Galactic Nuclei (AGN) in the high-z Universe
\citep{Harikane23b, Maiolino24}, like the so-called Little Red Dot
population \citep{Matthee24, Kocevski25, Akins25} suggest that AGN
contamination may play a role in boosting the UV luminosities for
composite sources. In addition, the treatment of dust attenuation
could also play a role, e.g. if outflows are able to effectively
remove dust from the galactic environment \citep{Fiore23,
  Ziparo23}. From a purely theoretical point of view, several
plausible modifications of the treatment of the baryon cycle at high-z
have been proposed, including an enhanced efficiency of star formation
and/or inefficient stellar feedback \citep{Dekel23, Li24,
  Cantarella25}. Moreover, an extremely bursty star formation history
in the first galaxies would lead to enhancement of the UV flux, with
respect to the expectation of models based on smooth star formation
histories \citep{Gelli24, Semenov25}.

Another possible explanation for the UV luminosities of high-z
galaxies could be a variation in the shape of the stellar initial mass
function (IMF \citealt{Trinca24, Yung24, Cueto24, Hutter25,
  Mauerhofer25}). Indeed, an IMF with an enhanced fraction of massive
stars with respect to the local measurements (hereafter a MW-like IMF
like a \citealt{Kroupa01} or \citealt{Chabrier03} IMF), would imply a
lower mass-to-light ratio and an enhanced production of UV photons per
unit star formation rate. On the other hand, it would also predict an
larger fraction of Type II supernovae and a lower fraction of mass
locked in long-living stars (per solar mass formed). Moreover, it is
important to keep in mind that assuming a variable IMF also implies
that current estimates for galaxy physical properties available in the
literature have to be revised \citep{Fontanot17a}, as the universality
of the IMF represents a strong prior often used in SED fitting codes.

Evidences for a different shape of the IMF with respect to the one
observed in the local neighbourhood have been growing in recent
times. In the local Universe, both dynamical, lensing and
spectroscopic properties of early-type galaxies (ETGs) show deviations
from the hypothesis of a universal MW-like IMF \citep{Cappellari12,
  vanDokkumConroy12, Dutton12, LaBarbera13}. In particular, dynamical
mass-to-light ratios are larger with respect to photometric estimates
assuming a universal IMF: this difference has been dubbed
$\alpha$-excess in the literature.  Moving to higher redshifts,
\citet{Gunawardhana11} found evidences for a flattening of the
high-mass end slope of the IMF at increasing star formation rate (SFR)
in the Galaxy And Mass Assembly (GAMA) survey. Moreover, several
results for early galaxies also favour a Top-Heavy IMF scenario for
JWST sources \citep{Katz22, Cameron24, Cullen25, Curti25}.

In our previous work \citep{Fontanot17a, Fontanot18a, Fontanot24}, we
have focused on the impact of several variable IMF scenarios on the
dynamical and spectroscopic properties of local ETGs, while also
discussing the impact of this hypothesis on their star formation and
galaxy assembly history. Our results showed a limited impact on the
reconstructed star formation histories of galaxies in different
stellar masses ranges. In particular, the trend for for shorter star
formation timescales in more massive galaxies is preserved. In the
framework of a variable IMF scenario, the main effect is connected
with the interpretation of the local chemical enrichment relations in
ETGs. In particular, we showed that, in this framework, the level of
$\alpha$-enhancement and the [$\alpha$/Fe] ratios do not trace the SFR
timescale, but the maximal level of SFR reached in the past \citep[see
  also \citealt{TinsleyLarson79}]{Fontanot17a}. Moreover, we showed
that a variable IMF can significantly affect stellar masses derived
from photometry (M$^{\rm phot}_\star$). These photometric masses may
differ from the intrinsic stellar mass predicted by theoretical
models, thereby biasing the recovery of the intrinsic galaxy stellar
mass function (GSMF) from photometric surveys \citep{Fontanot19}. This
IMF-driven effect on galaxy observables is expected to be stronger in
the early Universe, due to the younger ages of stellar populations and
shorter timescales sampled by observations.

The aim of the present work is to explore the effect of a variable IMF
scenario on the comparison between theoretical predictions and JWST
results, both in the observational space (UVLFs) and in the physical
space (GSMF). This paper is organized as follows. In
Section~\ref{sec:simsam} we introduce the latest realization of the
\gaea~model, while in Section~\ref{sec:varimf} we discuss the assumed
variable IMF scenario adopted in this paper. We then present our
results for the predicted UV luminosity function in
Section~\ref{sec:uvlf}; while in Section~\ref{sec:mstar} we discuss
the effect of the variable IMF scenario on the stellar mass estimates
for z>6 galaxies and compare with recent GSMF measurements. Finally in
Section~\ref{sec:final} we discuss the implications of our finding and
present our conclusions. All magnitudes quoted in this work have been
computed in the AB system \citep{OkeGunn83}.

\section{Theoretical models}
\label{sec:simsam}
\subsection{The GAlaxy Evolution and Assembly (\gaea) model}
In this work, we consider predictions from the latest version of the
GAlaxy Evolution and Assembly (\gaea) model \citep[see][and references
  therein]{DeLucia24}. \gaea~is a semi-analytic model (SAM), designed
to follow the growth of galaxy populations in a cosmological volume
and across a wide redshift range. These models assume that Dark Matter
Haloes (DMHs) represent the natural sites for galaxy formation, using
the baryons locked in their potential wells since early cosmic
epochs. SAMs start from a statistical description of the evolution of
DMHs population defined over a cosmological volume and unfold the
complex network of physical mechanisms acting on the baryonic gas
numerically solving a system of coupled differential
equations. Individual prescriptions typically employ approximated
parametrizations (which are usually derived from either empirical,
numerical or theoretical arguments) to describe the key dependencies
on galaxy physical properties. A critical difference with respect to
hydro-simulations lies in the lack of an explicit treatment for gas
dynamics: this represents an advantage for SAMs, as it drastically
reduces the computational demand of the method, thus allowing for a
large flexibility in exploring the effect of individual physical
processes on global galaxy evolution over large (cosmological) volume
and across a wide redshift range. However, this choice also represents
a major limitation for model predictions: in their standard
implementation, SAMs are indeed not able to track the spatial
distribution of the major baryonic components in the multi-phase gas.

The latest rendition of the \gaea~model has been recently presented in
\citet[see also \citealt{Xie24} and
  \citealt{Fontanot25}]{DeLucia24}. This version combines several
updated prescriptions at the same time, in particular:

\begin{itemize}
\item[$\star$]{a non-instantaneous model for the chemical enrichment
  of the inter-stellar medium (ISM) \citep{DeLucia14};}
  \item[$\star$]{a stellar feedback model based on the results of
    high-resolution numerical simulations \citep{Hirschmann16};}
  \item[$\star$]{an AGN feedback scheme based on the modelling of cold
    gas accretion onto SMBHs \citep{Fontanot20};}
  \item[$\star$]{an explicit partition of cold gas into atomic and
    molecular component \citep{Xie17};}
\item[$\star$]{a detailed treatment for dynamical processes acting on
  the gaseous component of satellite galaxies \citep{Xie20}.}
 \end{itemize}

With respect to the realization discussed in \citet{DeLucia24}, we
modify the prescription for stellar feedback, following results from
\citet{Cantarella25}. In detail, \gaea~uses a parametrization proposed
by \citet{Muratov15} using numerical experiments up to z$\sim$4 and
extrapolating the relevant trends to higher redshifts (see Eq.~14
and~15 in \citealt{Hirschmann16}). In the present work we assume that
at z$\gtrsim$4, stellar feedback efficiencies are capped at their z=4
values. \citet{Cantarella25} showed that the weaker stellar feedback
at z>4 leads to an increase of the normalization of the z$\sim$6 GSMF,
while having little impact on the bright end of the UVLF at
z$\gtrsim$10. We consider this model variant as our reference
\gaea~realization.

As in previous work, we use as a calibration set the evolution of the
GSMF up to z$\sim$3, the evolution of the AGN LF up to z$\sim$4 and
the local HI and H$_2$ mass functions. In a series of papers, we
compare model predictions against several properties of galaxy
populations across a wide range of cosmic epochs. In particular, we
conclude that our reference \gaea~run is able to correctly reproduce:

\begin{itemize} 
\item[$\star$]{the fractions and space densities of quiescent
  galaxies up to z$\sim$4 \citep{DeLucia24};}
\item[$\star$]{the emission line properties of different galaxy
  populations \citep{Scharre24};}
\item[$\star$]{the clustering properties of galaxy populations as a
  function of stellar mass, star formation activity, HI content and
  redshift \citep{Fontanot25};}
\item[$\star$]{the low-redshift dwarf galaxy (M$_\star \lesssim$ 10$^9
  \msun$) quiescent fractions \citep{Meng23};}
\item[$\star$]{the decreasing trend of HI and SFR of satellite
  galaxies in cluster halos \citep{Chen24};}
\item[$\star$]{the evolution of the UVLF and of the evolution of the
  mass-metallicity relation in the redshift range 5<z<9
  \citep{Cantarella25}.}
\end{itemize}

In \citet{Cantarella25}, we also considered predictions for the UVLF
and GSMF at redshift z$\gtrsim$9-10, finding an increasing tension
between \gaea~expectations and observational measurements. In detail,
we showed that this tension cannot be completely reduced by
considering the effect of the Eddington bias (at least considering up
to 0.5 dex uncertainty in the mass determination), and we discussed
alternative theoretical scenarios able to explain the mismatch, such
as a weaker stellar feedback or a feedback-free starburst scenarios
\citep{Dekel23}. In the present work, we will explore an alternative
scenario, which assumes that the stellar IMF associated with early
star formation episodes does not match the functional form derived in
the Milky Way (i.e. a \citealt{Kroupa01} or \citealt{Chabrier03} IMF),
but it favours the formation of massive stars (i.e. a so-called
``top-Heavy IMF'').

\subsection{The IMF scenario}
\label{sec:varimf}
\begin{figure}
  \centerline{ \includegraphics[width=9cm]{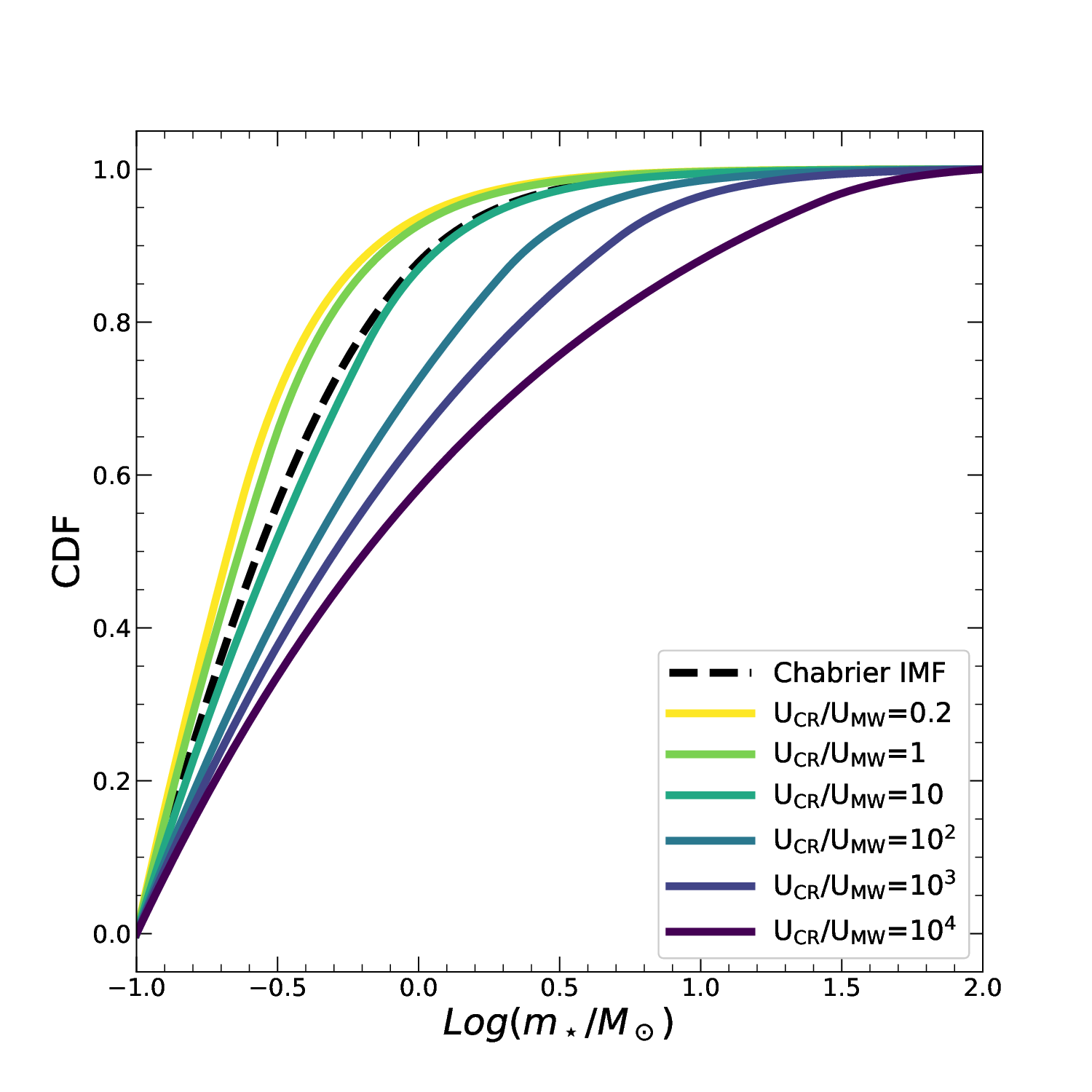}}
  \caption{Cumulative distribution function (CDF) for the different
    IMFs used in this work. Individual IMF shapes correspond to the
    IMF library described in \citet{Fontanot18a}, which assumes a
    dependence of the IMF characteristic stellar mass (i.e. the IMF
    knee) on the CR density background ($U_{\rm CR}$), following the
    numerical simulations of \citet{Papadopoulos11}. At increasing
    $U_{\rm CR}$, the characteristic mass get larger, which translates
    into an increasing the number of massive stars (i.e. a top-heavy
    IMF - see main text for more details). The black dashed line
    corresponds to a MW-like IMF, with the shape corresponding to a
    Chabrier IMF.}\label{fig:varimf}
\end{figure}
In our previous work \citep{Fontanot17a, Fontanot18a, Fontanot24}, we
explored the effect of a variable IMF on galaxy properties predicted
by \gaea. In particular, we focused on a subset of variable IMF
scenarios \citep{Kroupa13, Papadopoulos11, Weidner13, Fontanot18b},
linking the shape of the galaxy-wide IMF to the global SFR level (or
its density) of model galaxies. We showed that these runs correctly
reproduce a number of observational properties of low-z galaxies, that
are difficult to reconcile with a Universal, MW-like, IMF. One of the
critical predictions of the variable IMF framework implies that each
star formation episode can be characterized by a different IMF shape,
based on the physical properties of the interstellar medium of the
host galaxy. This also implies that the IMF varies with both cosmic
time and environment.

In this work, we focus on a single model to infer the impact of a
variable IMF scenario on the interpretation of high-z
observations. Specifically, we adopt the model used in
\citet[hereafter F18]{Fontanot18a}, which is based on the numerical
results of \citet[PP11 hereafter]{Papadopoulos11}. This framework
assumes that cosmic rays (CRs) associated with stellar evolution
(supernovae and stellar winds) can penetrate deeper into the inner
cores of molecular clouds (MCs) than UV photons, thereby influencing
the ionization and chemical state of star-forming regions. PP11
employed high-resolution simulations to show that the primary effect
of varying the CR background ($U_{\rm CR}$, normalized to the Milky
Way value $U_{\rm MW}$) is the modification of the expected Jeans mass
as a function of MC density. F18 used these numerical results to
generate a library of variable IMFs whose shapes scale with the SFR
density, assumed here to be a proxy for the CR field. In detail, we
assume that the evolution of the IMF shape is mainly driven by changes
in its characteristic mass, while the low- and high-mass slopes remain
fixed (see, e.g., Fig.~1 in F18). Overall, this implementation leads
to an increasingly top-heavy IMF with rising SFR density,
corresponding to a larger relative fraction of massive stars per unit
stellar mass formed; Fig.~\ref{fig:varimf} shows the cumulative
distribution function for the IMFs in the library (the corresponding
IMF shapes are shown in Fig.~1 - left panel - in
\citealt{Fontanot18a}). This choice represents an interesting scenario
given the increasing prevalence of compact galaxies and dense
starbursts at z > 6 in JWST observations \citep{AlvarezMarquez23,
  AlvarezMarquez25}.

\subsection{Estimating physical properties from synthetic photometry}
The main advantage of our approach is that, thanks to the availability
of simple stellar population (SSP) models associated with our IMF
shapes, we can compute self-consistent broadband photometry for the
resulting composite stellar populations and predict both luminosity
functions and stellar masses. In particular, we took advantage of
simple stellar population synthesis model libraries \citep{Bruzual03,
  Vazdekis10} built for the IMF shapes we consider. We then use these
to construct composite stellar populations combining stars formed with
the desired IMFs at different times, and use them to self-consistently
predict synthetic spectral energy distributions (SEDs) and broad band
photometry, corresponding to our model galaxies.

We use the spectro-photometric information associated to each model
galaxy to infer its physical properties, in a similar way to what an
observer would have derived\footnote{In our previous papers we
referred to these estimates as ``apparent'' physical properties
\citep[see e.g.][]{Fontanot17a}}, assuming a universal MW-like IMF. In
order to speed the calculations up while maintaining a classic SED
fitting approach, we use the 2-colour method introduced by
\citep{Zibetti09} to estimate mass-to-light ratios. A comprehensive
Monte Carlo library of composite stellar population synthesis models,
with a large variety of SFHs and chemical enrichment histories, is
used to infer the mass-to-light ratio in the rest-frame $i$ band
($M_\star/L_i$) as a function of the restframe (u-g) and (g-i)
colours. This library is bound to include only SFHs consistent with
the age of the Universe at z=6 and is based on the SSP models produced
by the \citet{Bruzual03} spectro-photometric code, assuming a Chabrier
IMF. In practice, we bin the models in the 2D (u-g) vs (g-i) colour
space and for each bin we consider the median $M_\star/L_i$ of the
corresponding models. We then estimate photometrically equivalent
stellar masses ($M^{\rm phot}_\star$) for each object in our sample
starting from its synthetic photometry and extracting the
$M_\star/L_i$ value from the closest position in this map. For the
small sample of sources with colours outside the reference map, we
extrapolate a $M_\star/L_i$ value via a nearest neighbour approach. We
include all relevant details about the generation of the $M_\star/L_i$
map in Appendix~\ref{app:ml_coltab}.


\subsection{Runs}
\begin{figure*}
  \centerline{ \includegraphics[width=18cm]{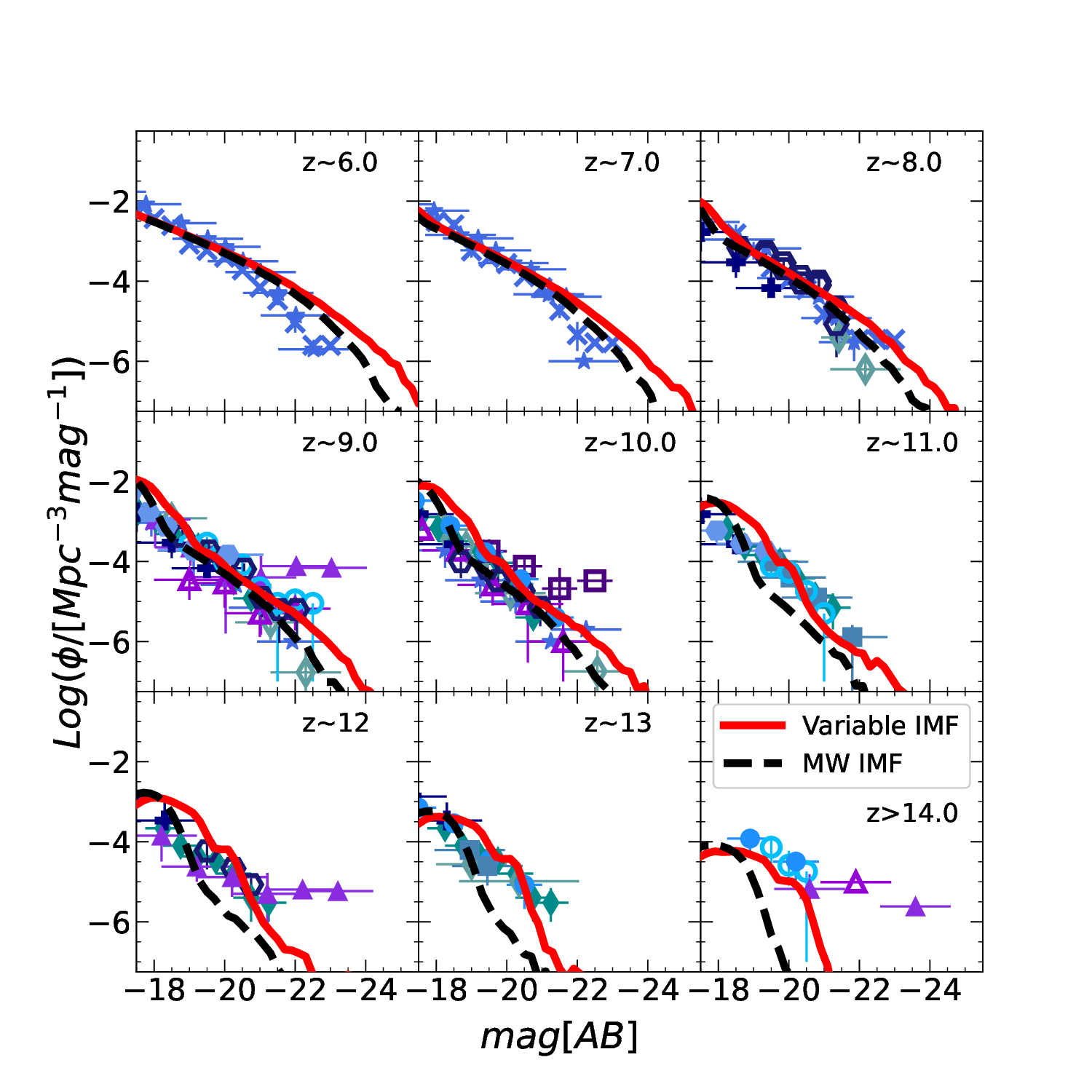}}
  \caption{Redshift Evolution of the UV Luminosity Function at
    z>6. Black dashed and red solid lines correspond to the
    predictions of the \gaea~realization assuming a universal MW-like
    IMF and the variable IMF scenario from
    Sec.~\ref{sec:varimf}. Observational datapoints from
    \citet[crosses]{Finkelstein15}, \citet[stars]{Bouwens21},
    \citet[pluses]{PerezGonzalez23}, \citet[empty
      squares]{Castellano23}, \citet[empty diamonds]{Donnan23},
    \citet[empty triangles]{Harikane23}, \citet[filled
      hexagons]{Leung23}, \citet[filled diamonds]{Donnan24},
    \citet[filled triangles]{Harikane24}, \citet[filled
      squares]{McLeod24}, \citet[empty circles]{Finkelstein24},
    \citet[empty hexagons]{Adams24}, \citet[filled
      circles]{Whitler25}. }\label{fig:uvlf}
\end{figure*}
In the following, we consider prediction draw from the latest
\gaea~realization presented in \citet[see also
  \citealt{Fontanot25}]{Cantarella25}, which couples the model with
merger trees extracted from the \pmill simulation (\citealt{Baugh19}
-- \pms hereafter). The \pms~ is a DM-only numerical simulation
defined over a 800$^3$ Mpc$^3$ volume and run using the {\sc gadget}
code \citep{Springel05} assuming cosmological parameters consistent
with the first year results from the Planck satellite
(\citealt{Planck_cosmpar} -- $\Omega_\Lambda=0.693$, $\Omega_m=0.307$,
$\Omega_b=0.04825$, $n=0.9611$, $\sigma_8=0.8288$, $H_0=67.77 \, {\rm
  km/s/Mpc}$). The particle mass resolution is $1.06 \times 10^8
\hmsun$ and merger trees have been constructed starting from DMHs
resolved with at least 20 particles, over a grid of 128 snapshots,
equally spaced in expansion factor\footnote{This correspond to roughly
half of the available snapshots in the \pms. This choice help reducing
the computational load and output size of the model, without affecting
the results as shown in \citep{Cantarella25}.}. This implies that
\gaea~reliably models galaxies down to a stellar mass limit of
$\sim$10$^8 \msun$. In \citet{Fontanot25} we showed that
\gaea~predictions are consistent for runs over other cosmological
simulations belonging to the same suite, assuming different
$\Lambda$CDM parameters (i.e.  the {\sc millennium} and {\sc
  millennium-II} simulations), that have been used in our previous
work.

In the following we consider predictions from two separate
\gaea~realizations. A first one corresponds to our standard run,
assuming a universal IMF in the form proposed by
\citet{Chabrier03}. We then run a new model, implementing the variable
IMF described in F18 and keeping the same model parameters as in the
standard run. As in F18, we assume that the SFR surface density
($\Sigma_{\rm SFR}$), computed using the disk radius (3.2 times the
exponential scale length of the gaseous component), is a good proxy
for $U_{\rm CR}$, such as:

\begin{equation}
\frac{U_{\rm CR}}{U_{\rm MW}} = \frac{\Sigma_{\rm SFR}}{\Sigma_{\rm MW}}
\end{equation}

\noindent
where $\Sigma_{\rm MW} = 10^{-3}$ M$_\odot$ yr$^{-1}$ kpc$^{-2}$,
represents the reference value for the MW disc. We do not attempt to
recalibrate\footnote{The model realization considered in F18 and
calibrated on low-z LFs was based on an earlier version of the
\gaea~model.} the run with a variable IMF to best reproduce the
available information on the evolution of the UVLF, which can be then
considered a direct predictions of our framework, highlighting the
impact of a variable IMF hypothesis on the predicted photometric and
physical properties of model galaxies. For both realizations, we saved
the physical properties for all galaxies in the \pms volume, plus we
compute the observed-frame photometry in each of the 8 JWST NIRcam
wide filters, the restframe UV luminosity using a top-hat ﬁlter
centred at 160 nm and 20 nm wide and the restframe optical photometry
in the ugi-SDSS filters.

\section{Results}
\begin{figure}
  \centerline{ \includegraphics[width=9cm]{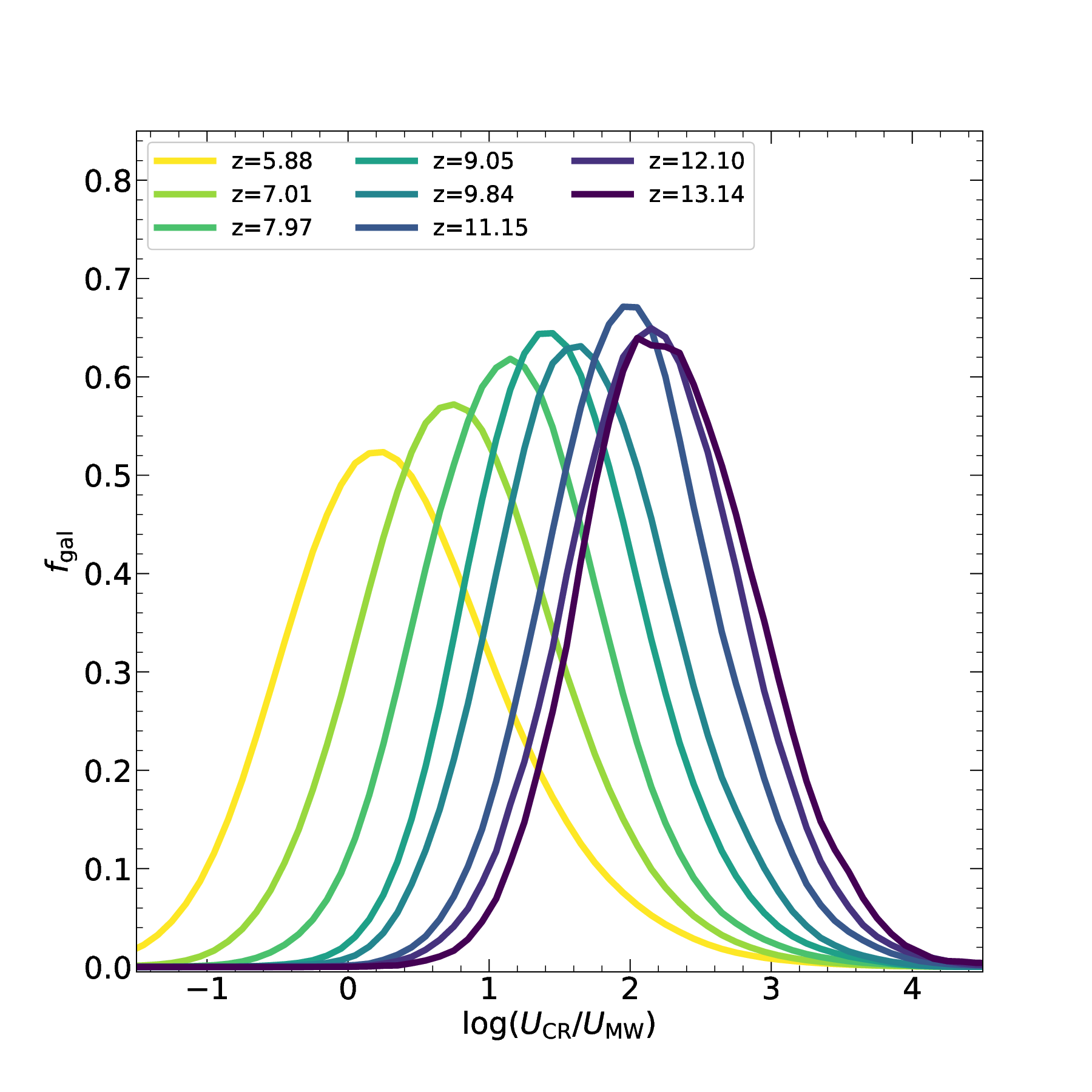}}
  \caption{Predicted $U_{\rm CR}/U_{\rm MW}$ distribution for
    \gaea~model galaxies in the variable IMF realization. Different
    redshift are shown in different colours as in the
    legend.}\label{fig:ucr}
\end{figure}
\begin{figure}
  \centerline{ 
    \includegraphics[width=9cm]{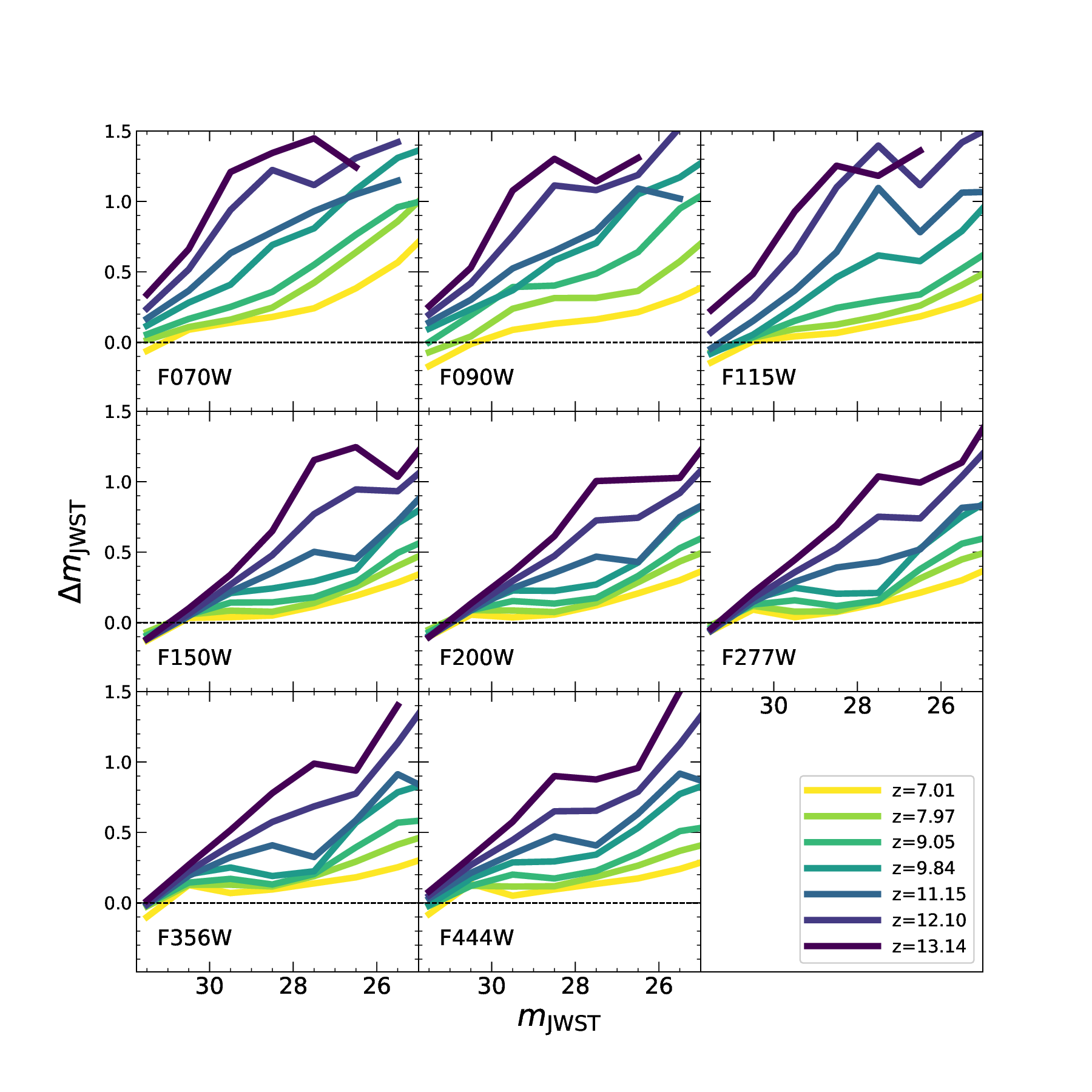}}
  \caption{Photometric differences in JWST filters between a
    \gaea~realization assuming a universal MW-like IMF and the run
    with the variable IMF. Each panel shows results for a given JWST
    wide filter (indicated in the label), predictions at different
    redshifts are marked by a different colour as in the legend. 
  }\label{fig:deltaMag}
\end{figure}
\begin{figure}
  \centerline{ \includegraphics[width=9cm]{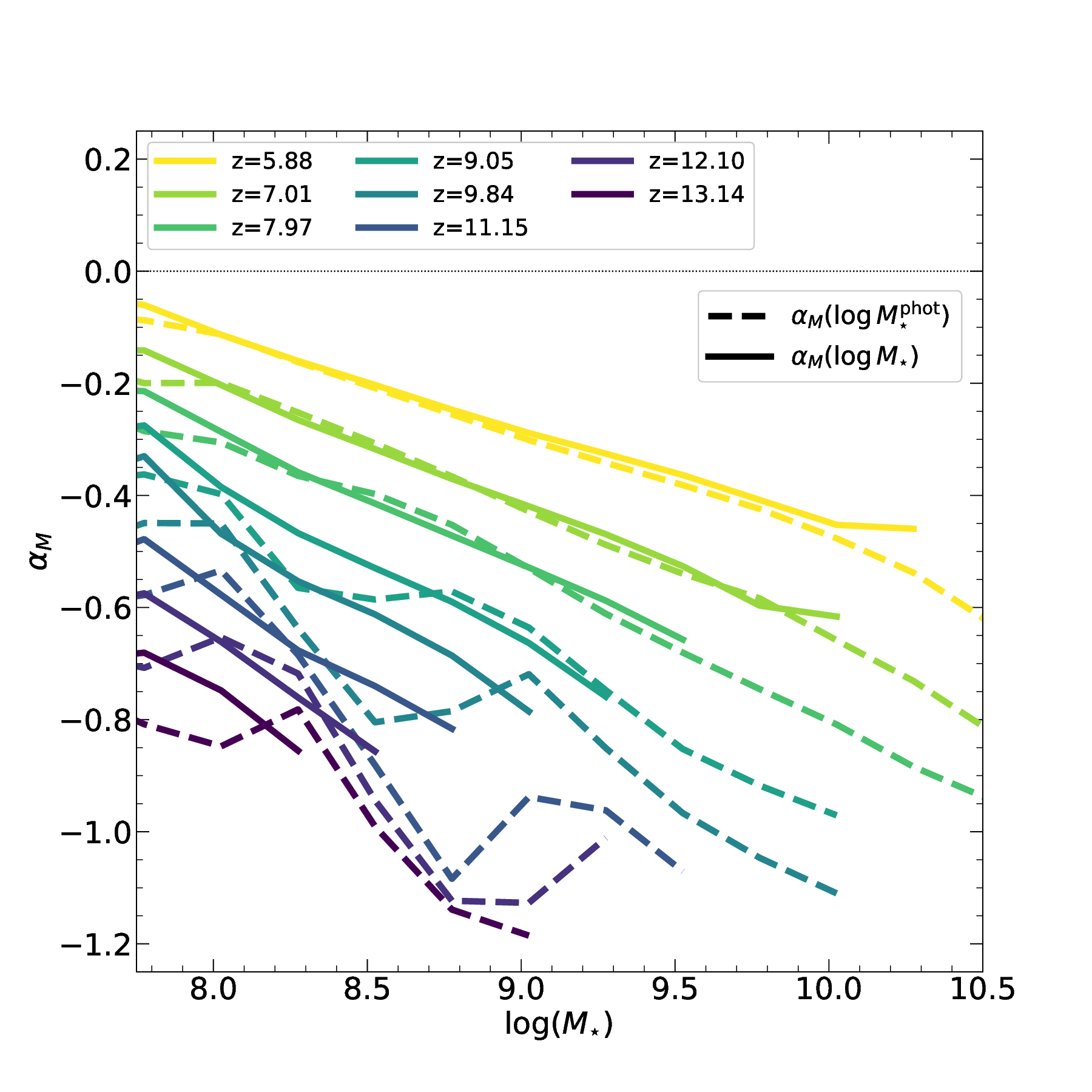}}
  \caption{Logarithmic Mass difference \alfex between the intrinsic
    stellar mass in the variable IMF run and M$^{\rm phot}_\star$,
    i.e. the stellar mass reconstructed from synthetic photometry
    assuming a MW-like IMF. Coloured lines refer to different
    redshifts as in the legend. Solid and dashed lines refer to the
    \alfex dependence on M$_\star$ and M$^{\rm phot}_\star$,
    respectively. Note that, on average, stellar masses in the
    variable IMF scenario are lower than those estimated for a MW-like
    IMF (see the main text for a discussion)}\label{fig:deltaMass}
\end{figure}
\subsection{Luminosity Functions}
\label{sec:uvlf}
We first focus on the predicted galaxy UVLFs from our
\gaea~realizations. In Fig.~\ref{fig:uvlf}, we compare the predictions
from our standard run (i.e. \citealt{Cantarella25}, black dashed
line), with the run with a variable IMF (red solid line). In both
cases we show predictions corresponding to unattenuated magnitudes,
i.e we are not considering the effect of dust. This choice is driven
by the consideration that the standard treatment for dust attenuation
in \gaea, is based on a simple model calibrated to low-z observations
\citep{CharlotFall00}, and might not represent an ideal framework to
model dust attenuation at higher redshifts. Moreover,
\citet{Cantarella25} showed that the \gaea~unattenuated UVLFs are in
better agreement with observations at z$\gtrsim$7-8. In a forthcoming
paper, we plan to better characterize the impact of dust extinction in
\gaea, by including an explicit modelling for the growth and
destruction of dust grains \citep{Osman26}.

Overall, the variable IMF scenario leads to an increase of the space
density of UV sources at all magnitudes above the resolution of the
model. The effect is larger at increasing redshifts, which is a
consequence of the larger contribution of top-heavy IMFs: in detail,
the variable IMF run predictions (red lines) are consistent with most
of the observed LFs and they converge to the standard run predictions
at z$\lesssim$7 (black dashed lines). This convergence in the
predictions at decreasing redshift suggests a dominant role of compact
star forming galaxies at high-z, corresponding to a larger
contribution of stellar populations associated with a top-heavy
IMF. This is clearly shown in Fig.~\ref{fig:ucr}, that displays the
predicted $U_{\rm CR}/U_{\rm MW}$ distribution for our model galaxies
at different redshifts. There is a clear evolution of the peak of the
distribution with redshift, which reflects in a shift of the typical
IMF associated with star forming galaxies at different
redshifts. Higher-redshift sources tend to be characterized by
top-heavier IMFs in the PP11 variable IMF \gaea~realization.

In our approach, different IMF shape are associated with different SSP
libraries. We use this information to generate synthetic photometry
for each model galaxy, that self-consistently takes into account the
IMF variations along the its formation history. For each source in our
simulated volume, we can compare the observer-frame magnitudes in the
9 wide JWST filters both for the reference run, assuming a MW-like IMF
(m$^{\rm MW}_{\rm JWST}$) and in the variable IMF run (m$_{\rm
  JWST}$). In Fig.~\ref{fig:deltaMag}, we show the mean difference
($\Delta m_{\rm JWST}=$m$^{\rm MW}_{\rm JWST}-$m$_{\rm JWST}$) between
these two predictions. The magnitude difference is not only driven by
the different SSPs, but also incorporates the changes in the physical
properties of high-z stellar populations, due to the IMF variations,
i.e. the returned fraction, the chemical enrichment and the fraction
of baryons locked into low-mass stars. Indeed, the star formation
histories of individual model galaxies are different between the two
runs. In Fig.~\ref{fig:deltaMag} we show the magnitude difference
between the variable-IMF and the standard-IMF predictions as a
function of m$_{\rm JWST}$. Each panel is for a single JWST filter
(blue to red, starting from the top left panel), while different lines
refer to different redshift (as indicated in the legend). Overall,
magnitudes in the variable IMF realization are brighter than the
corresponding estimates in a universal IMF run. The difference tend to
increase with galaxy luminosity and redshift, consistent with the
results from the UVLF analysis. At magnitudes around 30, differences
are negligible at most redshift but for the bluest filters, while they
can reach up to 1.5 magnitudes at 26 mag for the most extremes IMF
variations. These results suggest that IMF variations might offer a
plausible theoretical explanation for the extremely luminous sources
observed at the highest redshifts, without requiring changes to the
star-formation or feedback prescriptions relative to those adopted at
lower redshift \citep[see e.g.][and references
  herein]{Cantarella25}. Fig.~\ref{fig:deltaMag} shows that the
variable IMF run goes in the right direction and can account for a
large part of the tension between observations and the predictions of
the reference model. Next we analyse whether this can provide also a
quantitative solution to these tensions.
\begin{figure*}
  \centerline{ \includegraphics[width=18cm]{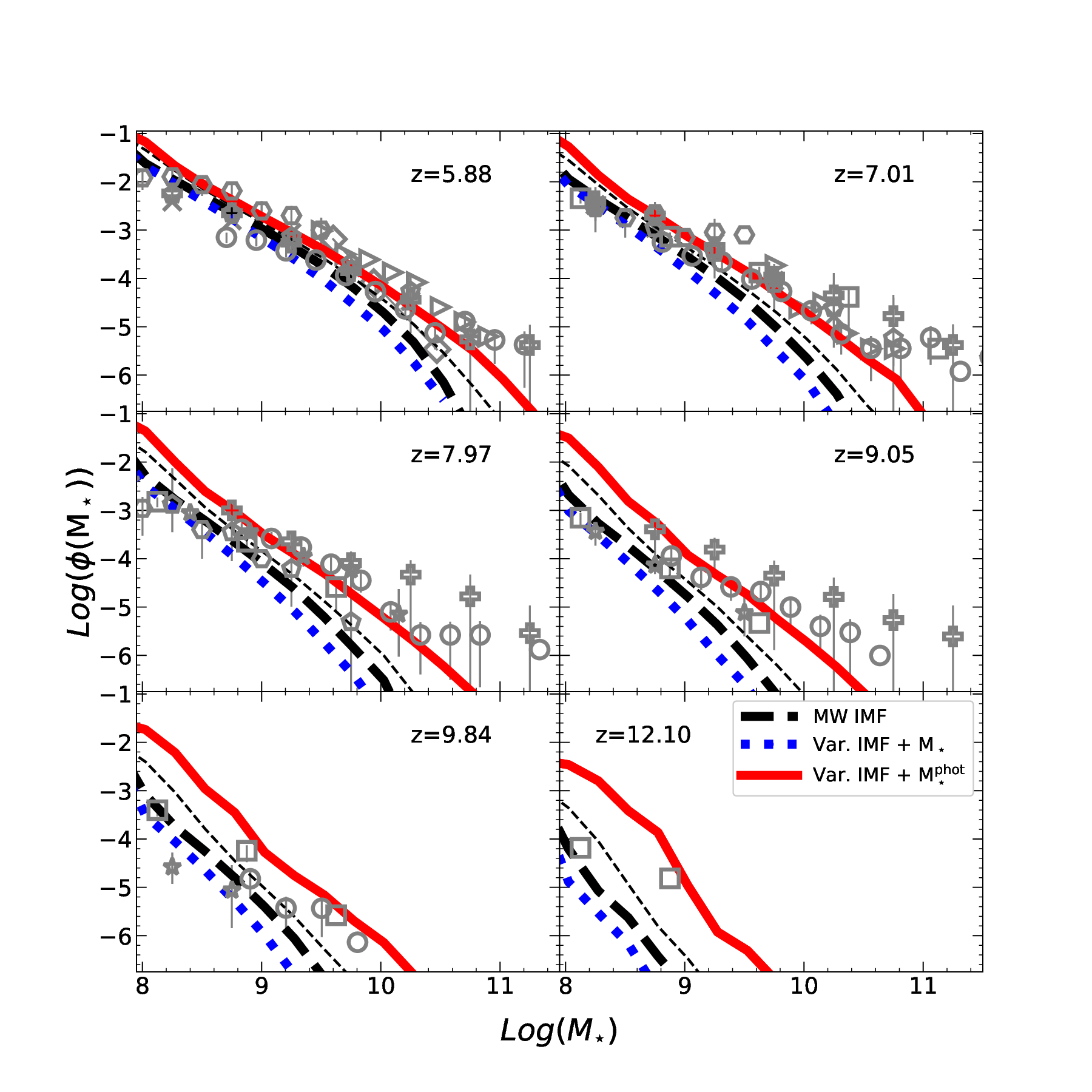}}
  \caption{Evolution of the galaxy stellar mass function at
    z>6. Dashed black, blue dotted and red solid lines refer to
    \gaea~predictions for the standard - MW-like IMF - run, the
    variable IMF run using the intrinsic M$_\star$ and the variable
    IMF realization using the photometrically derived M$^{\rm
      phot}_\star$. Thick and thin lines refer to intrinsic GSMF and
    to realizations taking the Eddington bias into account (see text
    for more details). Predictions are compared with data from
    \citet[crosses]{Gonzalez11}, \citet[triangles]{Grazian15},
    \citet[pentagons]{Song16}, \citet[diamonds]{Stefanon15},
    \citet[stars]{Stefanon21}, \citet[hexagons]{NavarroCarrera24},
    \citet[pluses]{Weibel24}, \citet[circles]{Shuntov25},
    \citet[squares]{Harvey25}.  }\label{fig:gsmf}
\end{figure*}

\subsection{Estimate for stellar masses}
\label{sec:mstar}
Taking advantage of the \gaea~coupling with spectro-photometric codes,
we can use the self-consistently derived photometry for each of our
model galaxies to provide an estimate for the expected stellar masses
assuming a MW-like IMF. It is worth stressing than since our code is
predicting JWST wide filter magnitudes, it would be possible to run
exactly the same pipelines that have been employed in JWST surveys on
both realizations. 

As we discussed in the introduction, a number of observational results
at low-redshifts reported a mismatch between the estimate of the
stellar mass of a galaxy derived from dynamical modelling and/or from
the analysis of spectroscopic features and the corresponding
photometric estimate (obtained through mass-to-light argument or via
SED fitting algorithms, assuming a MW-like IMF). This mismatch has
been referred to as the $\alpha$-excess. In our previous theoretical
work \citep[starting from][]{Fontanot17a}, we assumed that the
intrinsic stellar mass predicted by the model correspond to the
dynamical stellar mass, while we derived the photometric stellar mass
from synthetic photometry. We thus define an $\alpha$-excess for model
galaxies assuming \alfex=log(M$_{\star}$/M$^{\rm phot}_{\star}$), and
we study its dependence on both redshift and stellar mass.
Photometric stellar mass estimates can suffer from systematic biases
beyond IMF mismatch, particularly from assumptions about the SFH. To
quantify these, we compute \alfex for galaxies in the MW-like IMF
realization — i.e., matching the assumed SSP library IMF. At 6<z<8,
photometric estimates recover the intrinsic M$_{\star}$ to within a
few 0.01 dex, but at higher redshifts we find negative alpha excess
reaching $\sim$0.2 dex at z$\sim$12. Two effects drive this. First,
the SSP library includes models up to the Universe's age at z=6, which
exceeds the maximum age at higher redshifts. These models therefore
contain larger fractions of low-mass, low-luminosity stars that
inflate the mass-to-light ratio. Second, galaxies at higher redshift
are likely to have intrinsically burstier SFHs, which tend to have
systematically lower mass-to-light rations than smoother SFHs. As
shown by \citet{GallazziBell09}, broad-band colours alone cannot
distinguish SFH types; therefore, the use of an SSP model library not
specifically designed to match bursty SFHs leads to systematic
M$_{\star}$ overestimates of up to 0.1–0.2 dex. While confirming that
photometric masses may be significantly biased, this test places an
upper limit of ~0.2 dex on SFH-driven systematics.

In Fig.~\ref{fig:deltaMass}, we show the resulting $\alpha$-excess for
our model galaxies in the variable IMF run. We consider the \alfex
dependence on both M$_{\star}$ (solid lines) and M$^{\rm
  phot}_{\star}$ (dashed lines). It is quite clear that the absolute
value of \alfex strongly increases with redshift, thus reflecting the
increasing contribution of stellar populations characterized by
top-heavy IMFs. Fig.~\ref{fig:deltaMass} also highlights two important
properties of the model galaxy populations in the variable IMF
realization. First, the \alfex at z$\gtrsim$7-12 is both larger than
the SFH-driven systematics we already discussed and larger than the
claimed typical errors on derived M$_{\star}$ at these redshifts
(i.e. a statical error of the order of 0.1 dex of the reconstruction,
see e.g. \citealt{Carnall23}): this result stresses that a variable
IMF scenario may have a relevant impact in the interpretation of
observed data. Second, at variance with lower-redshifts ETGs (which
are characterized by a positive $\alpha$-excess), the recovered \alfex
systematically lie at negative values. This is due to the fact that at
low-z ETGs have been selected to be passive galaxies dominated by old
stellar populations, while our high-z galaxy sample consists of star
forming sources dominated by young and intermediate aged stars. The
different composition of stellar populations induces the different
trends in \alfex: indeed, at young ages, the mass and luminosity
budget is dominated by massive stars, having a lower $M_\star/L_i$,
while at low redshifts, the mass is dominated by stellar remnants,
implying an \alfex>1.

These results have a direct impact on the estimated evolution of the
GSMF at z>6. In Fig.~\ref{fig:gsmf} we compare different theoretical
estimates for the GSMF against the available observational
determinations (points with errorbars, all derived assuming a MW-like
IMF). We consider the GSMFs based on three different estimates for
stellar mass: the M$_{\star}$ predicted from the \gaea~run assuming a
MW-like IMF \citep[black dashed lines - see also][]{Cantarella25}; the
actual M$_{\star}$ predicted by the run implementing the variable IMF
(blue dotted lines); the M$^{\rm phot}_{\star}$ derived from our
synthetic photometry, assuming a universal IMF to estimate the
mass-to-light ratio (red solid lines). Thick lines display the GSMFs
based on the models' intrinsic predictions, for all cases. In
addition, with thin lines we show the predictions of the GSMFs for the
MW-like IMF realization including the eﬀect of the observational
errors on M$_{\star}$, which we compute by convolving individual
stellar mass estimates with a lognormal error distribution with a 0.25
dex amplitude. In general, we see that by including errors, the
agreement of the predictions for this run with the observed datapoints
improves, due to the broadening of the GSMF at the high-mass end.

The model assuming a MW-like IMF (black dashed lines) provides a good
agreement with the data up to M$_\star$<10$^{9} \msun$ and z$\sim$9,
but it falls short at higher masses and redshifts. GSMFs estimates
from the variable IMF \gaea~run based on M$^{\rm phot}_{\star}$ (red
solid lines) show larger number densities over the whole stellar mass
range, in good agreement with the observational constraints. At lower
redshifts, (6<z<9), GSMFs are closer to the expectations for the run
with a MW-like IMF: this is consistent with the results from the F18
model, that shows good agreement between the z<3 GSMFs based on
photometric stellar masses and the predictions from \gaea~runs based
on the universal IMF. It is important to also consider the intrinsic
GSMF predictions for the \gaea~realization with a variable IMF (blue
dotted lines): these systematically lie below the black dashed
lines. This is due to the fact that a top-heavy IMF implies less
baryons to be locked in long living stars and a larger gas mass
fraction to be recycled back into the ICM and IGM to foster subsequent
stellar populations. The comparison between the blue dotted and the
red solid lines clearly shows the impact of a possible mismatch
between the IMF used to estimate physical quantities for high-z
sources and their intrinsic IMF when interpreting the properties and
evolution of high-z galaxy populations.

\subsection{Mass-Metallicity relation}
\label{sec:mrz}
\begin{figure}
  \centerline{ \includegraphics[width=9cm]{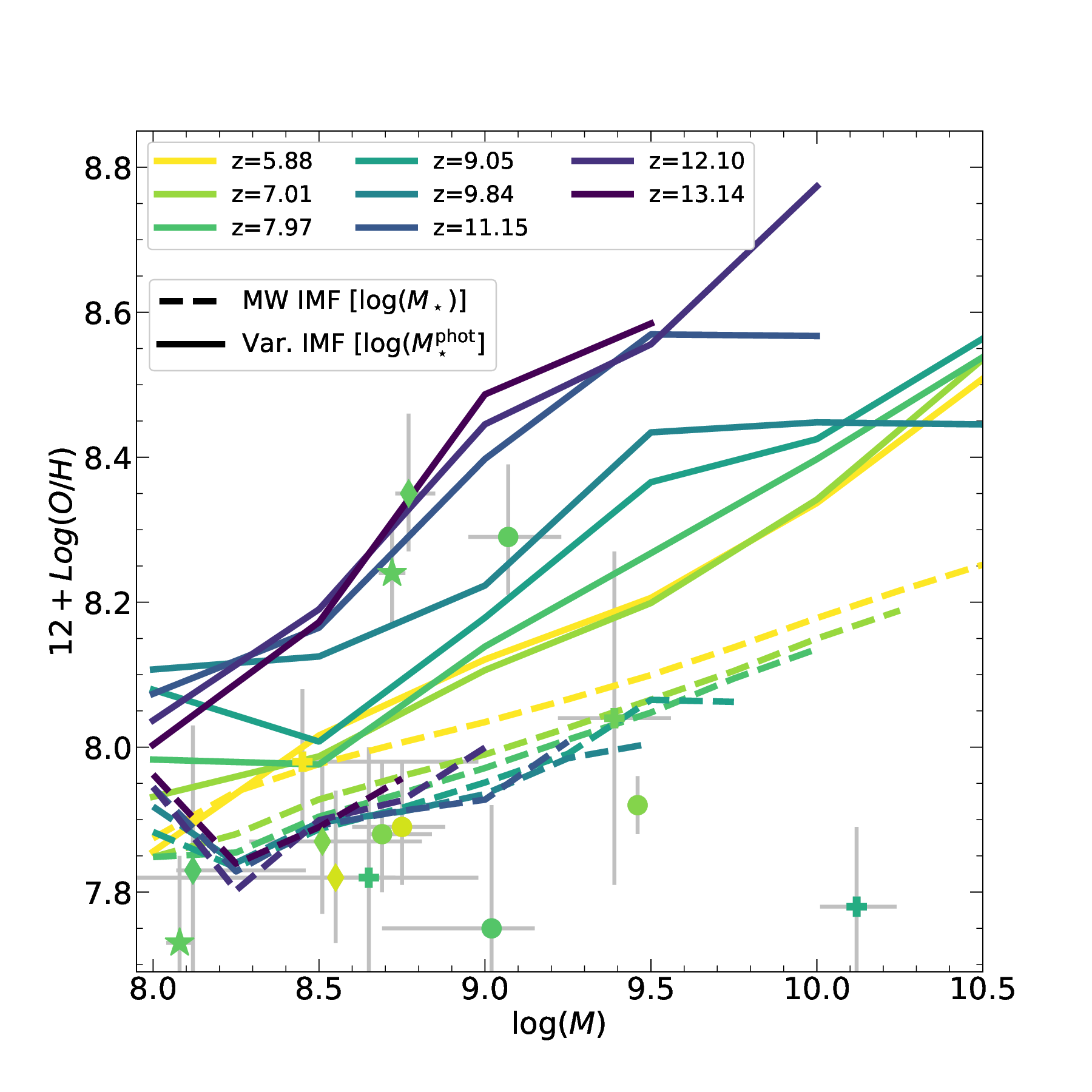}}
  \caption{Evolution of the gas phase mass-metallicity relation at
    z>6. Dashed lines refer to the predictions for the
    \gaea~realization assuming a MW-like IMF as a function of
    M$_\star$, while the solid lines correspond to the metallicities
    as a function of M$^{\rm phot}_\star$ in the \gaea~run
    implementing the variable IMF. Coloured lines refer to different
    redshift as in the legend. Datapoints from
    \citet[diamonds]{Nakajima23}, \citet[circles]{Morishita24} and
    \citet[stars]{Curti24}.  }\label{fig:mzr}
\end{figure}
An additional observational signature that is sensitive to the IMF is
metallicity. In Fig.~\ref{fig:mzr}, we show the redshift evolution of
the gas-phase MZR in the \gaea~run assuming either universal MW-like
IMF (dashed lines) or the variable IMF scenario (solid lines). In
order to improve the comparison with observational data, in the former
scenario the MZR is shown as a function of M$_\star$, while in the
latter as a function of M$^{\rm phot}_\star$. As in our previous work
\citep{Fontanot21, Cantarella25}, we consider only model galaxies with
a gas fraction larger than 0.1 and we apply a 0.2 downwards shift to
the predictions to account for the normalization shift between
theoretical predictions and the observed relation at z<0.2. In
\citet{Cantarella25}, we showed that model galaxies in our reference
run assuming a MW-like IMF are characterized by high gas-phase
metallicities: the resulting gas-phase MZR is offset high by $\sim$0.3
dex at 3$\le$z$\le$9, while the predicted slope of the relation is
consistent with the best-fit to the data
\citep{Morishita24}. Moreover, the overall MZR is hardly evolving in
shape and normalization at z>5.

An interesting prediction of the variable IMF realization is the
decreasing normalization of the MZR with redshift (of the order of
$\sim$0.3 dex from z$\sim$12-13 to z$\sim$6), which represent a trend
opposite to the reference run with MW-like IMF. In fact, in the
standard scenario with a time-invariant universal IMF, the overall
metal content is gradually increasing with cosmic time, as the result
of subsequent stellar populations, each of them releasing the same
metal yield in the ISM. On the other hand the variable IMF run is
characterized by top-heavier IMFs, implying larger SNe fraction and
larger metal enrichment at higher redshift. As redshift decreases, the
metal output from younger stellar populations decreases as well (since
they are born within an IMF with a decreasing SN fraction). Moreover,
the accretion of gas with pristine composition contributes to the
dilution of the overall ISM metallicity. Overall, the variable IMF
realization predicts gas-phase metallicities that are in tentative
agreement with some sparse JWST results at z$\gtrsim$6, based on the
electron temperature (T$_e$) method \citep{Nakajima23, Curti23,
  Curti24, Morishita24, Sanders24}. It is worth stressing, that most
of these determinations for high redshift galaxies use strong emission
(auroral) line. However, these method have been usually calibrated at
lower redshift: therefore they could result in systematic biases when
applied to high-z galaxies, due to their different ISM ionization
properties \citep[see e.g.][]{Sanders23a, Curti24}. Moreover, we are
assuming standard yields for short-lived stellar populations
(i.e. \citealt{ChieffiLimongi02}), usually calibrated on regular
Population II stars. If instead massive stars at the highest redshift
are characterized by different yields (e.g. Pop III stars
\citealt{Nomoto05}), this will likely have an impact on our
predictions. We will explore the impact of early Pop III enrichment on
the evolution of the MZR in future work (Cantarella et al., in preparation).


\subsection{Comparison with other models}

The results we present in this section can be compared with the
predictions of alternative theoretical models that attempt to
implement a variable IMF prescription. \citet{Yung24} explore the
impact of a top-heavy IMF within the framework of the Santa Cruz SAM
\citep{Somerville15}. Their approach does not include a
self-consistent implementation of a variable IMF prescription, but it
considers the increase in UV luminosity due to the expected increase
in the number of massive stars. They show that their model predictions
can be reconciled with observational determinations of the UVLF
assuming an increase in UV luminosity by a factor of 4, well within
the capabilities of a variable IMF scenario. Our results confirm their
suggestion by means of a self-consistent implementation of a variable IMF.

In a recent paper, \citet{Somerville25} propose a new variant of the
Santa Cruz SAM, implementing a density modulated star formation
efficiency. In this approach, motivated by results from numerical
simulations, star formation efficiency increases as a function of the
MCs surface density. \citet{Somerville25} show that this assumption
leads to $z>9$ space densities up to two orders of magnitude larger
than their reference model based on standard Schmidt-Kennicutt
prescription, in better agreement with the observational
constraints. This model employs a MW-like IMF, nonetheless it is
interesting to notice that, in order to reproduce the high-z JWST
observations, it resorts to a variation in the SFR prescription based
on the gas density. This assumption is conceptually consistent with
our approach of changing the IMF shape as a function of SFR density,
suggesting gas density as one of the main drivers of galaxy evolution
\citep[see also][]{Zibazzi22}.

The theoretical approach that is closer to ours is probably the one
implemented in the {\sc delphi} \citep{Mauerhofer25} and {\sc astreus}
\citep{Cueto24, Hutter25} suites. All these works implement a variable
IMF prescription inspired by the numerical results of \citet{Chon24}:
starting from a Kroupa functional shape, their IMFs become
progressively top-heavier, by assuming a log-flat distribution at the
massive end with a slope getting shallower at increasing redshift and
decreasing metallicity. They also compute a self-consistent synthetic
spectro-photometry for their model galaxies, by using stellar tracks
from Starburst99 stellar population synthesis code
\citep{Leitherer99}.

Their results confirm that the assumption of a variable IMF provides a
better fit to UVLFs at z$\gtrsim$10, thanks to the larger fraction of
massive stars, leading to brighter sources at fixed SFR.
\citet{Mauerhofer25} also consider the evolution of the GSMF, and they
show that a model with a variable IMF implies lower mass-to-light
ratios leading to smaller stellar masses at fixed UV luminosity. Their
intrinsic GSMFs underpredict the observational constraints at most
redshift. However, they do not attempt to infer the stellar masses
directly from their synthetic photometry under the assumption of a
MW-like IMF, so a direct comparison with literature data is
difficult. Both \citet{Cueto24} and \citet{Mauerhofer25} also consider
the evolution of the high-z gas-phase MZR: they also find an increase
in the normalization of the relation at increasing redshift
(corresponding to an increase of the top-heaviness of the relation),
consistent with our findings. However, the difference between the MZR
in their variable IMF scenario, with respect to runs with constant IMF
amounts to roughly 0.5 dex at all redshifts, while our model instead
predicts a strong redshift evolution of the MZR relation.

It is not easy to identify the physical origin of such different
trends. In \citet{Mauerhofer25} all {\sc delphi} model variants have
been independently calibrated to reproduce the 5<z<15 UV-LF by
construction, while in our approach we retain the same parameters
proposed by \citet{Fontanot25} to reproduce the GSMF and AGN-LF at
z<4. Our choice is further corroborated by the evidence that most
predictions from this specific variable IMF model converge to the
predictions of the standard run at z$\sim$6, thus ensuring that the
good agreement with the lower redshift galaxy populations is
preserved. Moreover, \gaea~also include both SMBH accretion and cold
gas partition prescriptions, which have been shown to be fundamental
to control the baryon cycle and the SFR \citep{DeLucia24}. A physical
mechanisms which is still lacking in the \gaea~framework, and most
likely of relevance for interpreting the high-z galaxy population is a
self consistent treatment of dust production and destruction. We
recently present such an implementation \citep{Osman26} and we are
planning to exploit it in the context of high-z galaxies and variable
IMF scenarios in a future work (Osman et al. in preparation).

\section{Summary and conclusions}
\label{sec:final}

In this work, we present the predictions of a \gaea~model implementing
a variable IMF scenario. We assume that the evolution of the IMF is
characterized by an increase of the characteristic stellar mass
(i.e. the knee of the IMF) at increasing SFR density, inspired by the
work of \citet{Papadopoulos11}. We stress that in the original work
the IMF variability is driven by the local CR density field, while in
this work we assume the SFR density as a proxy for CR density. In this
paper, we show that the impact of the variable IMF gets larger at
increasing redshift, with larger deviations from the properties
predicted by the reference run assuming a MW-like IMF at increasing
redshift. This result suggests a scenario where the global CR density
field increases with redshift, as a consequence of increasing density
and/or specific star formation rate. In our framework, this correspond
to a scenario with an increasingly top-heavier IMFs at increasing
redshift. As such, this paper represents an exploratory work to assess
the overall impact of the IMF variability on our interpretation of
high-z data, in particular from JWST surveys.

Our results show that stellar populations associated with a
top-heavier IMF, naturally predict an increase in the rest-frame UV
flux, with respect to a universal MW-like IMF scenario, driven by the
corresponding larger fraction of massive stars. When compared with
available data, our model is thus able to better reproduce the UVLF up
to z$\sim$13. \gaea~also predicts that the brightening is not only
limited to the UV restframe, but it affects all magnitudes
corresponding to JWST wide filters (from F070W to F444W). This effect
scales with the magnitude and redshift and it is particularly relevant
($\sim$1 mag) at the bright end (mag$\lesssim$27). The larger fraction
of massive stars also implies a larger fraction of SNeII and a larger
baryon returned fraction. Overall, this implies smaller stellar mass
for \gaea~model galaxies at fixed UV magnitude, with respect to the
standard realization. This corresponds to a negative $\alpha$-excess
(i.e. the ratio between the intrinsic stellar mass and the stellar
mass estimated under the assumption of a MW-like IMF - it has been
dubbed ``excess'' since this is typically positive at lower
redshifts), and GSMFs based on intrinsic stellar masses that have even
smaller space densities than the observational constraints.
Nonetheless, it is worth stressing that the model GSMFs for the
variable IMF realization built using the photometric stellar mass
estimates (i.e. the stellar masses derived from our synthetic
photometry assuming a MW-like IMF), are in good agreement with the
results from high-z surveys up to z$\sim$12. The variable IMF scenario
leaves also a strong imprint in the chemical enrichment history of
model galaxies. In particular it implies a normalization of the
gas-phase MZR at z$\gtrsim$6 larger than available estimates, a
steeper relation and a decreasing evolution of the normalization with
redshift (at variance with the model using a MW-like IMF).

All those results represent an attempt to self-consistently address
the effect of a variable IMF scenario in a galaxy formation model at
z>6. Consistently with this approach, we do not attempt any
recalibration of \gaea~model parameters with respect to the standard
run (assuming a Chabrier IMF): this implies that any different in
model predictions is only driven by the IMF variation. This in turn
implies that a more focused work on the variable IMF realization, may
also lead to a better agreement with the available data. We advocate
that the next step should involve more sophisticated models, self
consistently linking the IMF shape to host high-z galaxy properties
(see e.g. \citealt{Jerabkova18, Chon24, vanDokkumConroy24} among the
others).

A variable IMF scenario is also entangled with the study of a possible
contribution of the so-called Population III stars. In this work, we
consider IMF variations that only include Population II stars,
i.e. focusing on the relative fraction of stars in different mass
ranges. However, Pop III stars are expected to be characterized not
only by a different IMF \citep[see e.g.][]{Hirano15, Hosokawa16}, but
also by a different dynamical mass range (in most scenarios only
massive stars are expected to form, with typical masses much larger
than the maximum allowed for standard populations
\citealt{OmukaiPalla03}). Moreover, their chemical enrichment patterns
(i.e. stellar yields) are predicted to be quite different from MW-like
populations \citep[see e.g.][]{Nomoto05}. We leave a more detailed
study of the impact of a Population III scenario on the interpretation
of high-z observations in a forthcoming work (Cantarella et al., in
preparation).

\begin{acknowledgements}
  An introduction to \gaea, a list of our recent work, as well as
  datafile containing published model predictions, can be found at
  \url{https://sites.google.com/inaf.it/gaea/home}. We acknowledge the
  use of INAF-OATs computational resources within the framework of the
  CHIPP project \citep{Taffoni20} and the INAF PLEIADI program
  (\url{http://www.pleiadi.inaf.it}). FF acknowledges support from the
  INAF 2024 RSN1 Minigrant ``The eﬀect of a variable IMF on Galaxy
  Evolution and Assembly: from the local to the high-z Universe.'' and
  from the Next Generation European Union PRIN 2022 ``20225E4SY5 -
  From ProtoClusters to Clusters in one Gyr''. MH acknowledges funding
  from the Swiss National Science Foundation (SNSF) via a PRIMA grant
  PR00P2-193577 ‘From cosmic dawn to high noon: the role of BHs for
  young galaxies’.
\end{acknowledgements}

%
\bibliographystyle{aa} 
\bibliography{fontanot} 
%
\begin{appendix}\label{app:ml_coltab}
  
\section{Stellar masses from optical rest-frame photometry}
\begin{figure}[!t]
  \includegraphics[width=9cm]{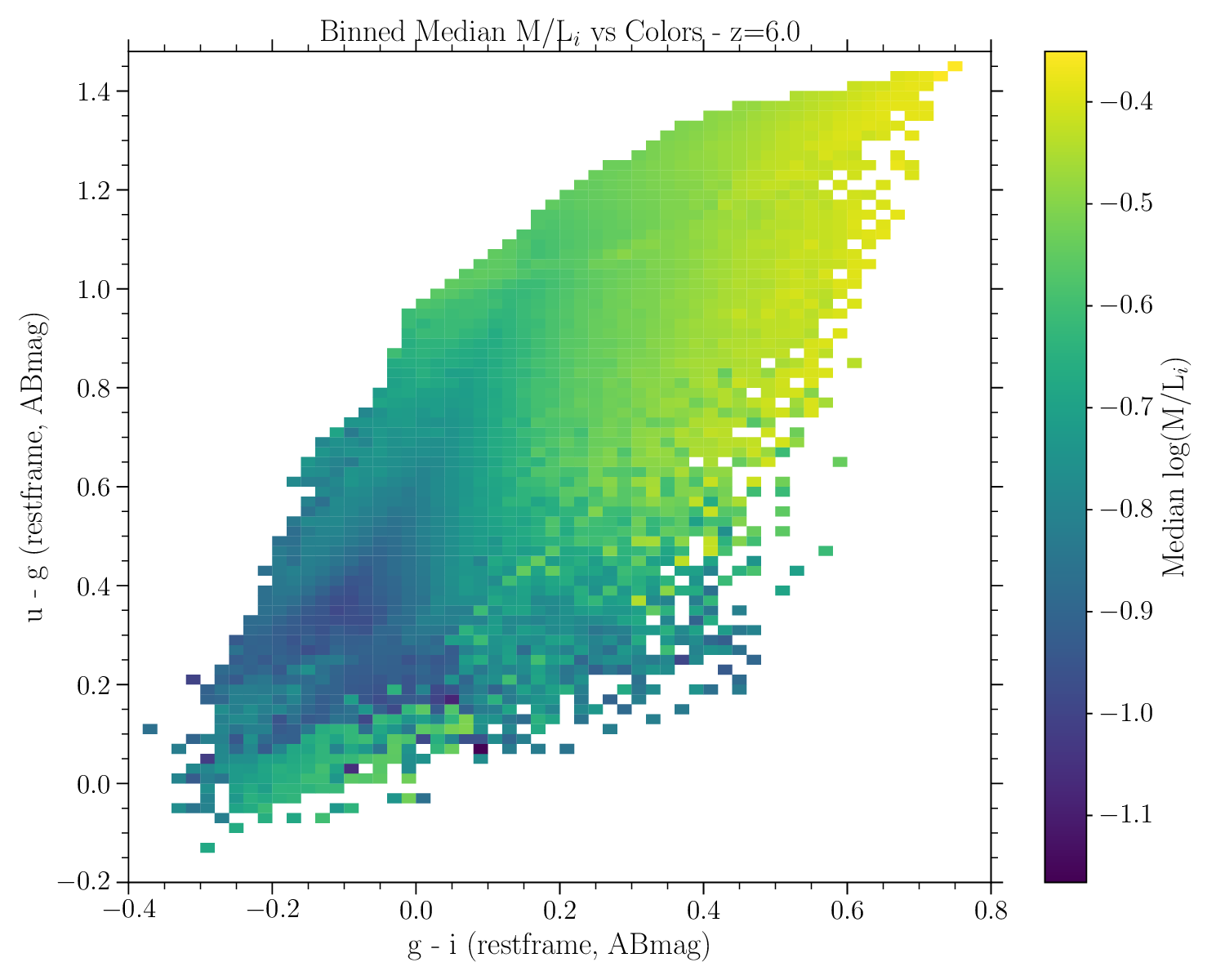}
  \caption{Stellar M/L in SDSS-$i$ band (solar units) in the
    colour-colour space defined by $u-g$ versus $g-i$ (rest-frame SDSS
    filters, AB magnitudes), as it results by assuming a standard
    \cite{Chabrier03} IMF and the modelling described in the appendix
    text. The map represents the median M/L of the models in bins
    $0.02$~mag $\times 0.02$~mag wide, according to the colorbar on
    the right. Only the 137,489 models with a SFH consistent with the
    age of the Universe at $z=6$, i.e. $t_\mathrm{form,lb}<0.92$~Gyr,
    are shown here. Note that the colours and the luminosities used for
    M/L computations are intrinsic, i.e. without any dust attenuation
    effect applied.}\label{fig:MLi_uggi}
\end{figure}

To simulate the stellar mass that an observer would infer from SED
modelling based on standard stellar population assumptions and, in
particular, a \cite{Chabrier03} IMF, we use the two-colour method
introduced by \cite[][ZCR09, hereafter]{Zibetti09}. The idea is to
build a vast library of composite SP models and express the M/L in a
given band as a function of two optical colours, which capture the SED
shape. Given a pair of ``observed'' colours from the simulations, the
average M/L of the models in the corresponding small 2D colour bin is
used to convert the observed luminosity into stellar mass.

Given the idealised nature of this experiment and to avoid folding
uncertainties concerning the dust treatment into the colour dependence,
we depart from ZCR09 by using models that include only pure stellar
emission. The derived colours are consistently compared with the
intrinsic stellar emission computed for the simulated galaxies.

The present library is based on the original SSP models of BC03, built
on STELIB stellar templates and Padova 1994 evolutionary tracks. This
choice ensures consistency with the SSPs used for the simulated
galaxies. The composite SPs are built according to the SFH and
chemical enrichment histories introduced by \cite{Zibetti17}. Each SFH
includes a secular continuous component modelled as a delayed Gaussian
function \`a la \cite{Sandage86}, characterized by a lookback time for
the beginning of the SFH ($t_\mathrm{form,lb}$) and a timescale $\tau$
that corresponds to the time elapsed until the peak of the SFR is
reached. These two parameters are randomly generated according to the
following priors. For $t_\mathrm{form,lb}$, we adopt a log-uniform
prior between 500 Myr and the age of the Universe at the redshift of
interest. Given $t_\mathrm{form,lb}$, $\tau$ is generated from a
log-uniform prior in $\tau/t_\mathrm{form,lb}$ between 1/50 and
2. Random bursts are added following the prescriptions of
\cite{Gallazzi05}. Notably, our models implement a simple chemical
evolution prescription, with a parameterization derived from leaky-box
models, plus a randomization for the bursts. The allowed range of
variation for the stellar metallicity is between 1/50 and 2.5
$\mathrm{Z_\odot}$ ($\mathrm{Z_\odot}\equiv 0.02$). More details about
the implementation of the bursts and of the chemical evolution are
provided in Appendix A of \cite{Mattolini25}.

Fig. \ref{fig:MLi_uggi} displays the variation of M/L in the SDSS $i$
band (in solar units) across the plane defined by the optical colours
$u-g$ versus $g-i$ (SDSS filters, AB system). Here we consider only
models with $t_\mathrm{form,lb}<0.92$~Gyr, corresponding to the age of
the Universe at $z=6$ (flat $\Lambda$CDM cosmology with $h=0.7$,
$\Omega_\Lambda=0.7$).

The 137,489 models are binned in $0.02$~mag $\times$ $0.02$~mag colour
bins, and the median M/L in each bin is calculated and represented in
the map. The resulting matrix is then linearly interpolated over the
entries with no models and extended using a nearest-neighbour algorithm
to the full colour-colour space. Each galaxy in the simulation is
assigned the M/L of the corresponding colour-colour bin in this
map. Note that the dispersion of M/L in each bin ranges between 0.05
and 0.15 dex, hence well below the typical mass excess determined in
Fig.~\ref{fig:deltaMass}.

Note that by further restricting the maximum allowed
$t_\mathrm{form,lb}$ for higher $z$, the number of models decreases
and the coverage of the space at the red end as well. However, we
verified that such a more restrictive $t_\mathrm{form,lb}$ selection
does not significantly affect the M/L in the common colour-colour
space. For this reason, we adopt the map for $z=6$ as the reference
look-up table for all redshifts.

\end{appendix}

\end{document}